\pdfoutput=1
\documentclass[journal,twoside,web]{ieeecolor}
\usepackage[pdftex]{graphicx}
\usepackage[outdir=./]{epstopdf}
\usepackage{generic}
\usepackage{cite}
\usepackage{amsmath,amssymb,amsfonts}
\usepackage{algorithmic}
\usepackage{graphicx}
\usepackage{textcomp}
\usepackage{systeme}
\usepackage{float}
\usepackage{verbatim}
\usepackage{dblfloatfix} 
\def\BibTeX{{\rm B\kern-.05em{\sc i\kern-.025em b}\kern-.08em
    T\kern-.1667em\lower.7ex\hbox{E}\kern-.125emX}}
\begin{document}
\title{Long Short-Term Memory Implementation Exploiting Passive RRAM Crossbar Array}
\author{Honey Nikam, Siddharth Satyam, and Shubham Sahay, \IEEEmembership{Member, IEEE}
\thanks{Honey Nikam and Siddharth Satyam are with the Department of Mechanical Engineering, Indian Institute of Technology Kanpur, Kanpur 208016, India (e-mail: honeyn@iitk.ac.in). }
\thanks{Shubham Sahay is with the Department of Electrical Engineering, Indian Institute of Technology Kanpur, Kanpur 208016, India (e-mail: ssahay@iitk.ac.in).}}
\maketitle

\begin{abstract}
The ever-increasing demand to extract temporal correlations across sequential data and perform context-based learning in this era of big data has led to the development of long short-term memory (LSTM) networks. Furthermore, there is an urgent need to perform these time-series data-dependent applications including speech/video processing and recognition, language modelling and translation, etc. on compact internet-of-things (IoT) edge devices with limited energy. To this end, in this work, for the first time, we propose an extremely area- and energy-efficient LSTM network implementation exploiting the passive resistive random access memory (RRAM) crossbar array. We developed a hardware-aware LSTM network simulation framework and performed an extensive analysis of the proposed LSTM implementation considering the non-ideal hardware artifacts such as spatial (device-to-device) and temporal variations, non-linearity, noise, etc. utilizing an experimentally calibrated comprehensive phenomenological model for passive RRAM crossbar array. Our results indicate that the proposed passive RRAM crossbar-based LSTM network implementation not only outperforms the prior digital and active 1T-1R crossbar-based LSTM implementations by more than three orders of magnitude in terms of area and two orders of magnitude in terms of training energy for identical network accuracy, but also exhibits robustness against spatial and temporal variations and noise, and a faster convergence rate. Our work may provide the incentive for experimental realization of LSTM networks on passive RRAM crossbar arrays.
\end{abstract}


\section{Introduction}
\label{sec:introduction}
\IEEEPARstart{T}{he} development of recurrent neural networks (RNNs) has led to the recent breakthroughs in the field of machine intelligence while analysing the patterns in the sequential/temporal data and performing efficient time series predictions for applications including speech and video processing and recognition, character-level language modelling and translation, ECG signal processing, image captioning, stock market modelling, air pollutant modelling, modelling active cases for COVID-19, etc. Unlike the feed-forward deep neural networks (DNNs), an RNN allows the past observations to be used as inputs for predicting the output at a later time instance [1]. However, the vanishing gradient problem limits the capability of RNNs in maintaining long-term dependencies [1]. To circumvent this issue, long short-term memory (LSTM) networks, a special class of RNNs, which are capable of learning long-term dependencies as the information is propagated along with the time flow were proposed [1]-[2].\par
However, the conventional implementation of LSTM networks on digital platforms with von-Neumann architecture such as general-purpose CPUs, or more advanced GPUs, or FPGAs incur a significantly high energy dissipation and a large inference latency due to the frequent exchange of data between the memory and the processor [3]-[8]. Since vector-by-matrix multiplication (VMM) is the prominent operation in LSTMs [1]-[2], their energy consumption can be reduced significantly by exploiting the massive parallelism and the inherent ability of the resistive random-access memory (RRAM) crossbar arrays to perform extremely energy-efficient VMM using the physical (Ohm's and Kirchoff's) laws [9]-[11]. However, the active 1T-1R crossbar arrays lead to a large area overhead since the select transistor has to sustain the forming/programming currents/voltages of the RRAM attached to their drain terminal, which are significantly high as compared to the typical operating currents/voltages of the advanced logic transistor technologies and restricts their incessant scaling. Considering the inherent scaling benefits of the passive RRAM crossbar arrays (which do not require a selector device) [12]-[14], it becomes imperative to explore their potential for implementation of compact and ultra-low power LSTM networks.
\par To this end, in this work, for the first time, we explore the possibility of performing the core computational tasks of the LSTM network i.e. vector-by-matrix multiplication (VMM) and in-situ training (using the Manhattan learning rule) on a passive RRAM crossbar array. We perform an extensive investigation of the impact of non-idealities of RRAM devices including spatial (device-to-device) and temporal variations, noise, etc. on the LSTM implementation utilizing an experimentally calibrated comprehensive phenomological model. Our results indicate that the passive RRAM crossbar array-based LSTM implementation converges to the optimal accuracy faster and exhibits a significant improvement in the area by $\sim$6.5x10$^3$ times and training energy by $\sim$51.7 times as compared to the active 1T-1R crossbar-based LSTM implementation [10]. Moreover, the memory density of the passive RRAM crossbars may be further upscaled with the help of 3D-integration [15] for accommodating the large number of LSTM parameters required for practical applications such as speech recognition and processing, language processing, video surveillance, etc.
\par The manuscript is organized as follows: sections I.A and I.B provide a brief introduction to the LSTM networks and the passive RRAM crossbar arrays, respectively. The developed LSTM simulation framework and the comprehensive compact model utilised for passive RRAM crossbar arrays in this work are discussed in section II. The performance metrics of the proposed passive RRAM crossbar-based LSTM network implementation such as accuracy, area and energy are highlighted in section III and conclusions are drawn in section IV.

\renewcommand{\thefigure}{1}
\begin{figure}[t]
    \centering
    \includegraphics[width=\linewidth]{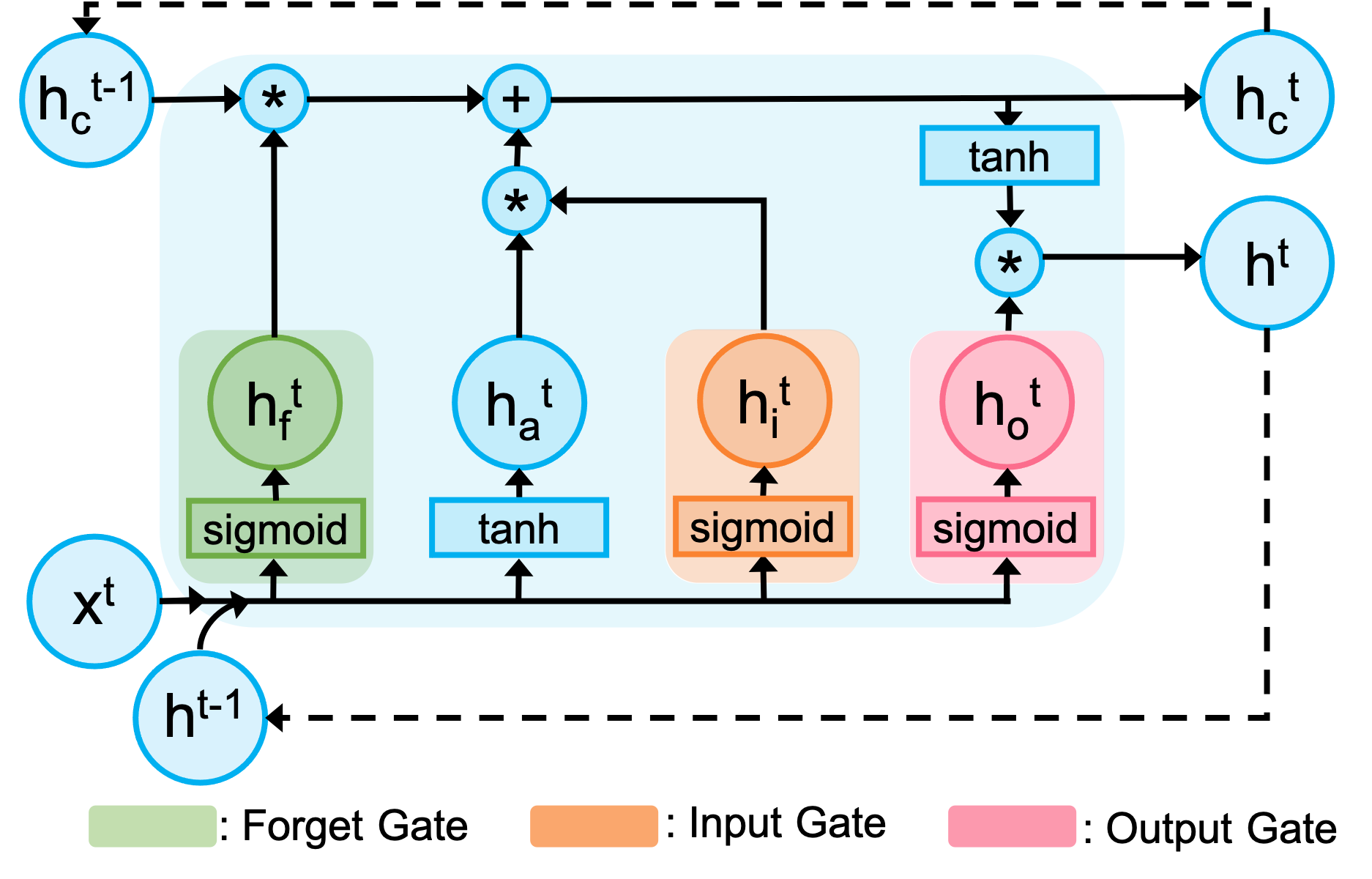}
    \vspace{-20pt}
    \caption{The schematic view of an LSTM cell.}
\vspace{-10pt}
\end{figure}

\subsection{Long Short-Term Memory Networks (LSTMs)}
RNNs may be perceived as DNNs with a feedback loop which allows the outputs at a previous time step to be used as inputs at a later time step. LSTM is a powerful variant of the RNNs suited for sequence prediction problems [1]-[2]. A typical LSTM consists of a repeating module or cell as shown in Fig. 1 that retains (and transmits) the temporal information in the form of a cell state $h_c^{t}$. The information carried by the cell-state can be modulated by the different gates (which are neural network themselves) in the LSTM cell. At any time step $t$, the LSTM cell is fed with a present input vector ($x^t$), and the hidden state ($h^{t-1}$) and the cell state ($h_c^{t-1}$) corresponding to the previous time step $(t-1)$. Each gate computes an output as a function of $x^t$, $h^{t-1}$, associated weights $W_f$, $W_a$, $W_i$, and $W_o$ and biases, $b_f$, $b_a$, $b_i$, $b_o$ corresponding to the forget gate, activation gate, input gate and output gate respectively. The forget gate facilitates removal of irrelevant pieces of information from the cell-state as denoted by equation (1):
\begin{align}
h_f^{t} &= \sigma(W_f \cdot [h^{t-1}, x^t] + b_f).
\end{align}

The input gate decides which pieces of information need to be updated in the cell-state based on the present input ($x^t$) and the previous hidden-state ($h^{t-1}$) and is governed by equations (2) and (3):
\begin{align} 
h_a^{t} &= \tanh{(W_a \cdot [h^{t-1}, x^t] + b_a)}\\
h_i^{t} &= \sigma(W_i \cdot [h^{t-1}, x^t] + b_i)
\end{align}

The cell-state of the LSTM is modified by first removing the irrelevant information via point-wise multiplication with the forget gate output and then performing point-wise addition of the input gate output before propagating to the next cell following equation (4):
\begin{align} 
h_c^{t} &= h_i^{t} \cdot h_a^{t} + h_f^{t} \cdot h_c^{t-1}
\end{align}

Now, based on the information provided by the present input ($x^t$), the previous hidden-state ($h^{t-1}$) and the modified cell state ($h_c^{t}$), the output gate determines the next hidden-state which is fed to the next cell following equations (5) and (6):
\begin{align} 
h_o^t &= \sigma(W_o \cdot [h^{t-1}, x^t] + b_o)\\
h^t &= h_o^t \cdot \tanh{(h_c^t)}.
\end{align}
Fig. 1 shows the typical flow of information through an LSTM network.
\renewcommand{\thefigure}{2}
\begin{figure}[t]
    \centering
    \includegraphics[width=\linewidth]{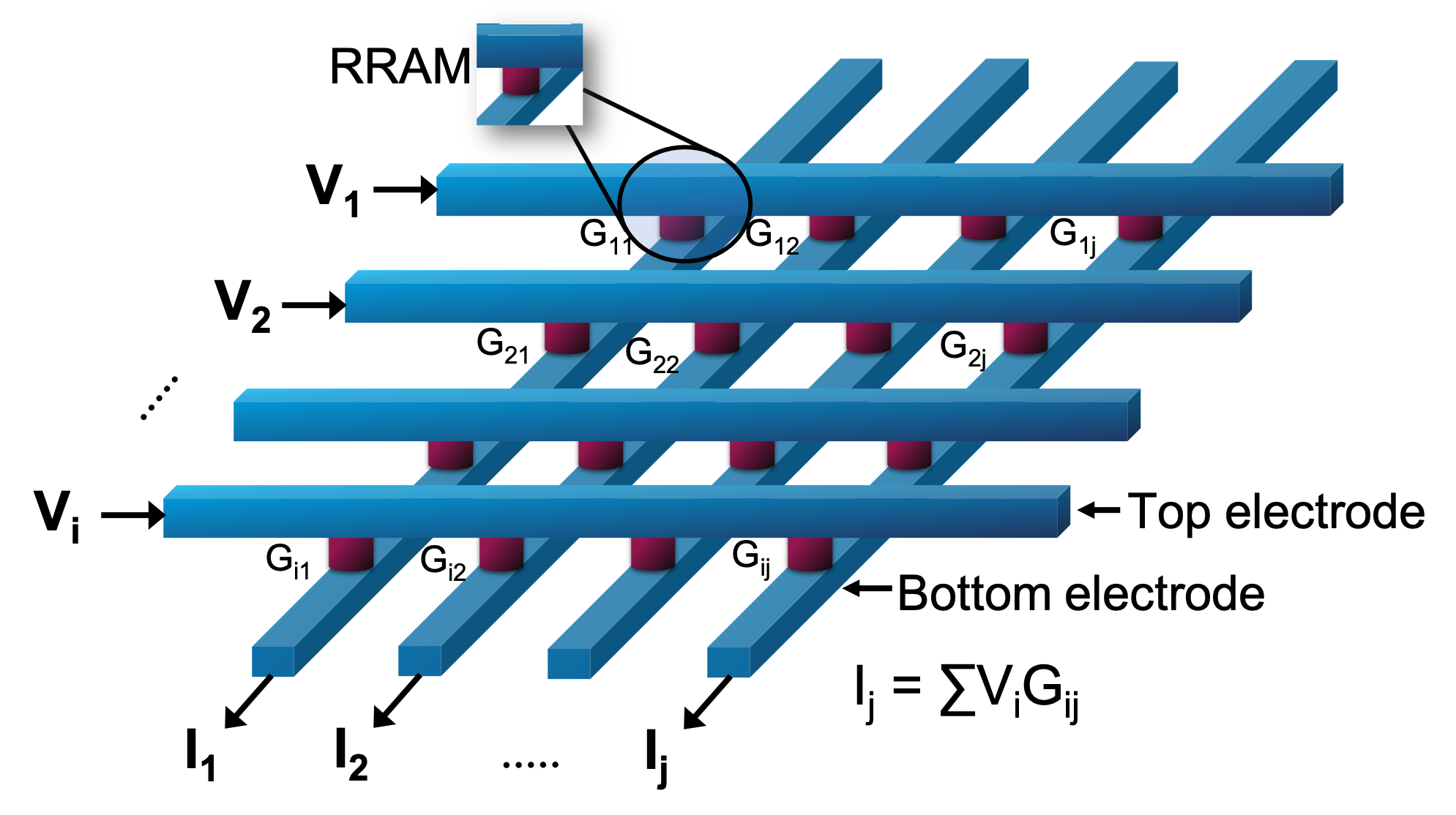}
    \vspace{-20pt}
    \caption{3D view of a passive RRAM crossbar array. RRAM devices arranged in a crossbar configuration perform efficient VMM. At each cross-point, the RRAM current is the product of voltage (input) and the RRAM conductance-state (weight), following Ohm’s law. The total current flowing through the column is the summation of the RRAM currents (VMM) following the Kirchhoff’s current law.}
\vspace{-10pt}
\end{figure}

\renewcommand{\thefigure}{3}
\begin{figure}[t]
    \centering
    \includegraphics[width=\linewidth]{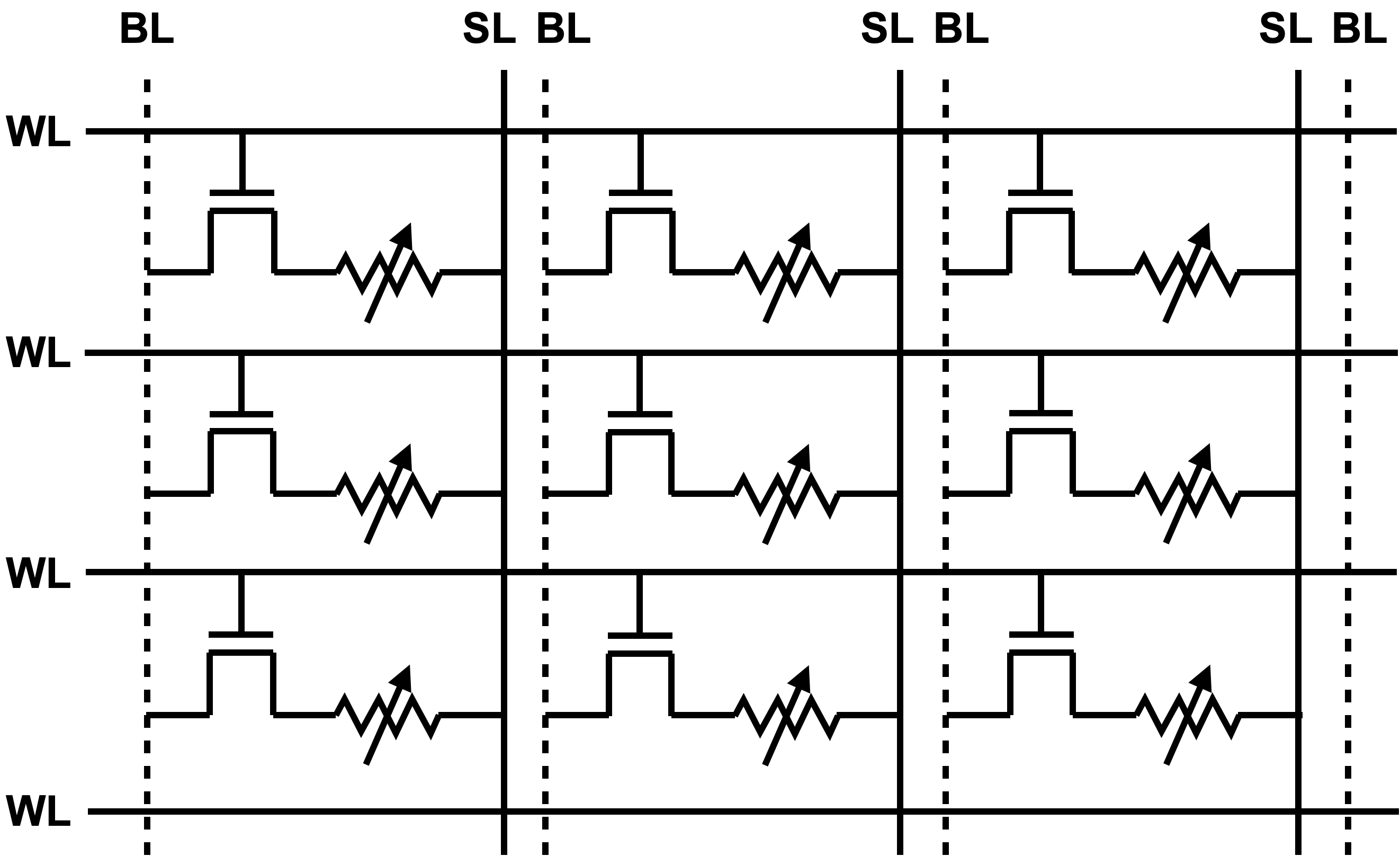}
    \vspace{-20pt}
    \caption{Schematic of the 1T1R active crossbar array}
\vspace{-10pt}
\end{figure}

\subsection{Passive RRAM crossbar arrays}
Resistive RAMs (RRAMs) are metal-insulator-metal (MIM) stacks in which an insulator layer (typically a transition metal oxide such as HfO$_x$, TiO$_x$, TaO$_x$, etc.) is sandwiched between two metal electrodes. Filamentary RRAMs not only exhibit a reversible resistive switching between two extreme resistance-states known as high-resistance state (HRS) and low-resistance state (LRS), but can also be tuned to any intermediate resistance-state with the aid of ultra-fast electrical pulses ($<$ 100 ns) [12]-[17]. RRAMs are available in two configurations: (a) active 1T-1R crossbar in which the RRAMs are embedded on the drain electrode of the select transistor and exhibit enhanced tuning accuracy due to precise control of RRAM current with the help of gate electrode of the select transistor and (b) passive crossbar array in which the RRAM devices are realised at the intersection of two perpendicular metal (top and bottom) electrodes as shown in Fig. 2.\par
The passive RRAM crossbars can be scaled incessantly to the ultimate scaling limits as they do not require a selector device for operation and can also be monolithically integrated in the back-end-of-line (BEOL) [12]-[17]. Although it is relatively difficult to precisely control the current through the RRAMs in a passive crossbar configuration due to absence of the selector device, prior studies have shown the possibility of tuning the resistance (or conductance)-state with a precision $>$ 7 bits using program-verify algorithm [17]. This unique conductance tuning capability allows RRAM crossbars to mimic the biological synapses and perform vector-by-matrix-multiplication in the analog domain if the inputs are encoded as voltages and weights are stored as conductance-states of the RRAMs using Ohm's law and Kirchoff's law [13]-[14] as shown in Fig. 2. The von-Neumann bottleneck is circumvented in such VMM implementations since the feed-forward propagation (VMM) is performed in-situ at the same location where the neural network parameters (weights) are stored [18]-[20]. Therefore, passive RRAM crossbars are promising candidates for realizing extremely area- and energy-efficient VMM engines.

\renewcommand{\thefigure}{4}
\begin{figure}[t]
    \centering
    \includegraphics[width=\linewidth]{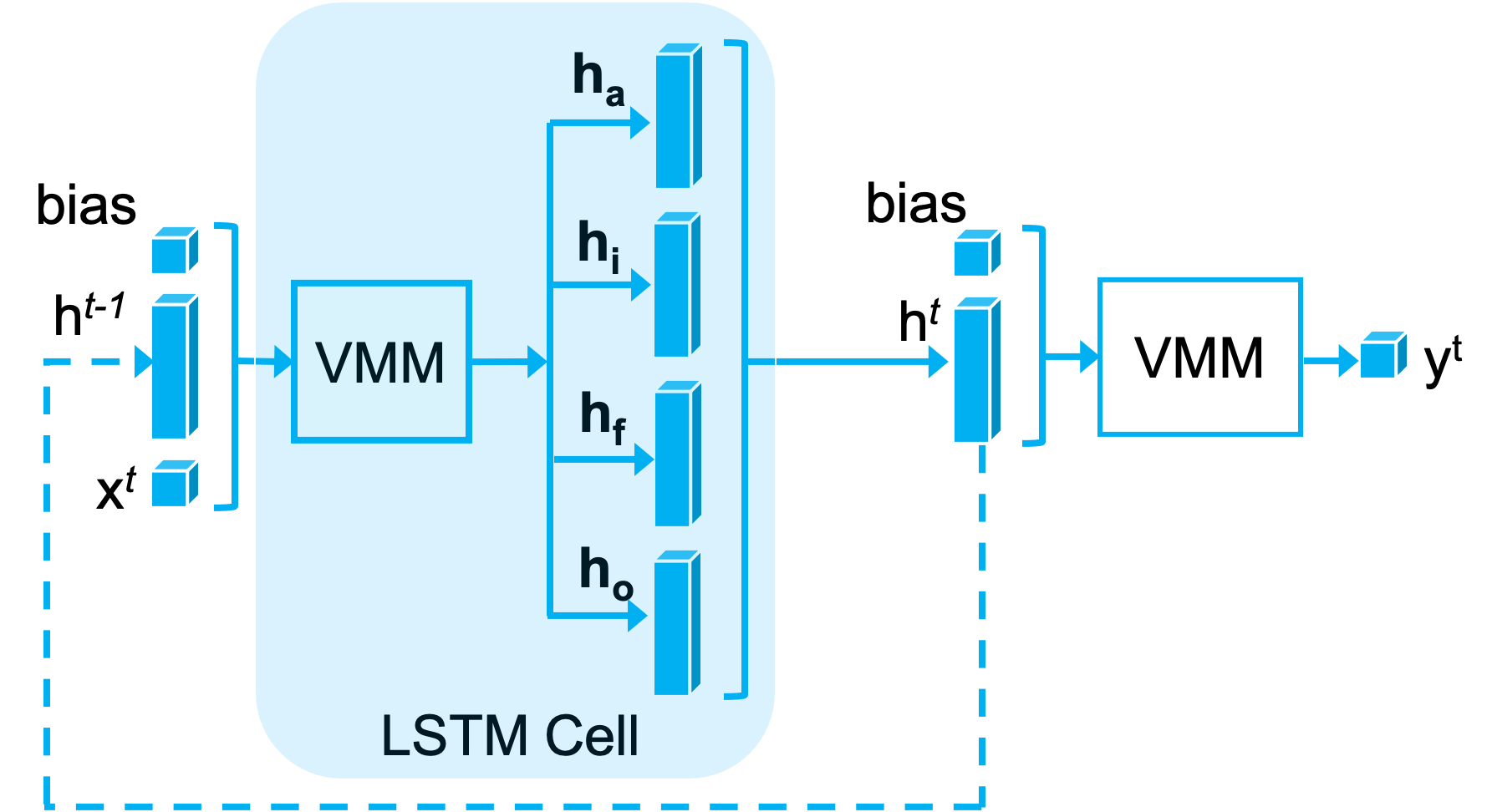}
    \caption{The RNN implemented using passive RRAM crossbar array in this work. It consists of an LSTM layer feeding a dense fully-connected layer.}
\vspace{-10pt}
\end{figure}

\section{Simulation Framework and Modeling Approach}
Considering the inherent scaling benefits and the efficacy of the passive RRAM crossbar arrays while performing energy-efficient VMM operation, we explored their potential for implementation of compact and ultra-low power LSTM-based RNNs. For proof of concept demonstration and a fair comparison, similar to [10], a multi-layer LSTM-based RNN was considered in this work with 15 LSTM units as shown in Fig. 4. Moreover, the output of the LSTM layer is fed to a dense fully-connected layer.
\par
We encoded the LSTM network parameters (weights) shared across the different time steps as the conductance-states of a passive RRAM crossbar array to perform in-situ computations. A comprehensive phenomological model for the static characteristics including noise and dynamic behavior considering spatial (device-to-device) and temporal variations of the passive RRAM crossbar array-based on the Pt/Al$_2$O$_3$/TiO$_{2-x}$/Ti/Pt stack [16] which is validated against more than 2 million experimentally characterized data points and 324 RRAMs has been utilised in this work. For extracting optimal performance from the proposed LSTM network implementation, the conductance range of RRAMs was restricted between 100 $\mu S$ to 300 $\mu S$ in this work based on the unique dynamic behavior of the passive RRAM crossbar array [16].\par
\renewcommand{\thefigure}{5}
\begin{figure}[t]
    \centering
    \includegraphics[width=\linewidth]{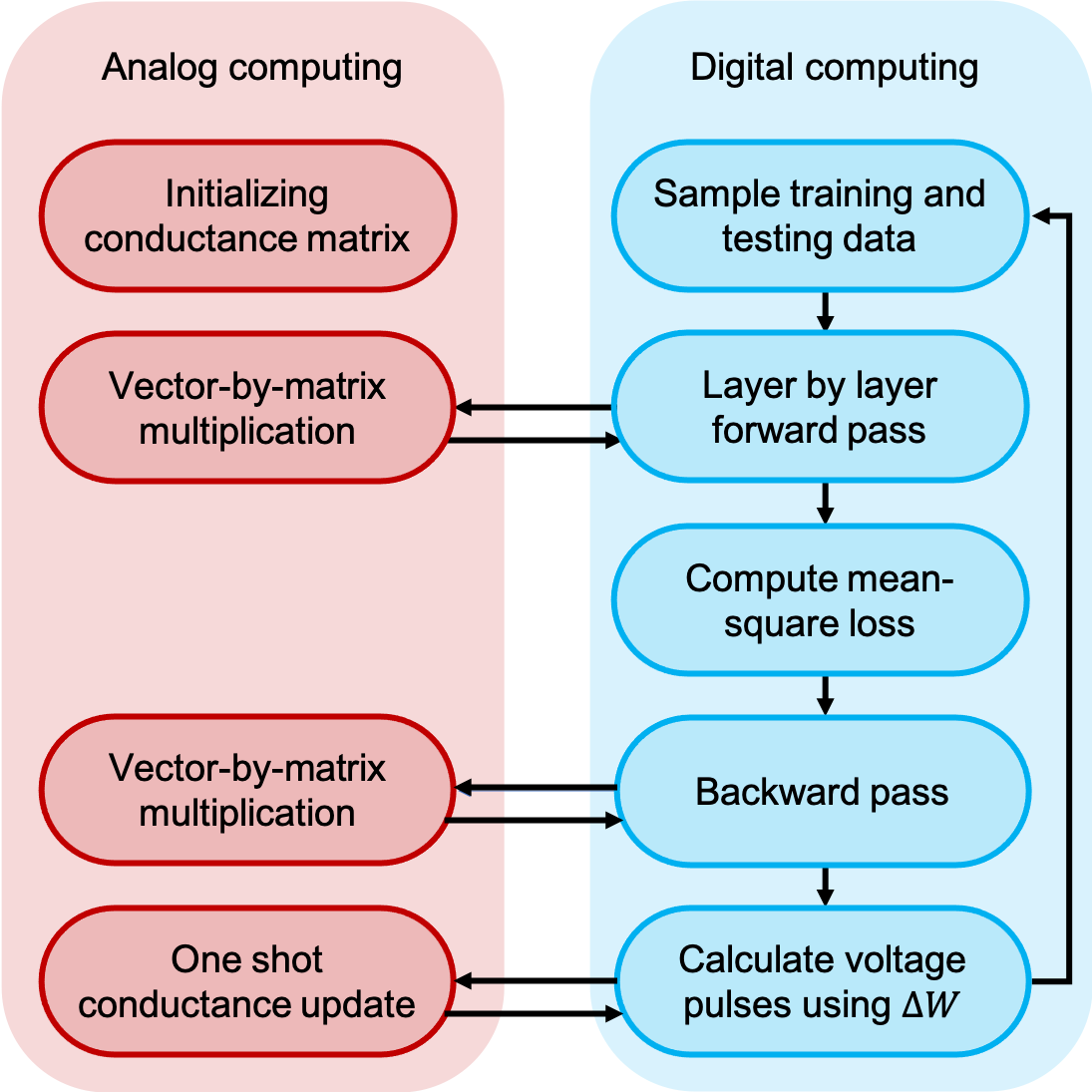}
    \caption{The simulation framework consists of a hybrid of computations performed in the digital domain and on the passive RRAM crossbar array (analog domain). Data sampling, forward pass, loss computation and backward pass were performed in the software digitally while the vector-by-matrix-multiplication and update of the conductance-states were performed on the passive RRAM crossbar array using the experimentally calibrated comprehensive compact model considering the hardware imperfections [16].}
    \vspace{-12pt}
\end{figure}
For the implementation of the recurrent-LSTM network, the ($64 \times 64$) passive RRAM crossbar array [12] was partitioned as follows: a $34 \times 60$ array stores the weight analogues for the LSTM layer while a $32 \times 1$ matrix stores the weight analogues for the fully-connected dense layer as differential pairs of conductance-states to account for both positive and negative weight values ($W_{ij}$ = $G^+_{ij}$ – $G_{ij}^{-}$). First, the conductance-states were randomly initialized between 100 $\mu S$ to 300 $\mu S$ representing random weights. Subsequently, the VMM during the layer-by-layer forward and backward pass and the weight (conductance-state) update during the training process utilising a hardware-friendly algorithm known as Manhattan rule [21] were performed in-situ on the passive RRAM crossbar array as shown in Fig. 5. Such in-situ computations outwit the von-Neumann bottleneck arising due to the frequent shuffling of information between the storage and processing units. Moreover, in-situ training with backpropagation is capable of self-adaptively adjusting the network parameters to fit the training data even in the presence of hardware imperfections such as noise, device-to-device variations, non-linearity, etc. [10]-[11], [13], [21].\par

\renewcommand{\thefigure}{6}
\begin{figure*}[!b]
    \centering
    \includegraphics[width=0.95\textwidth]{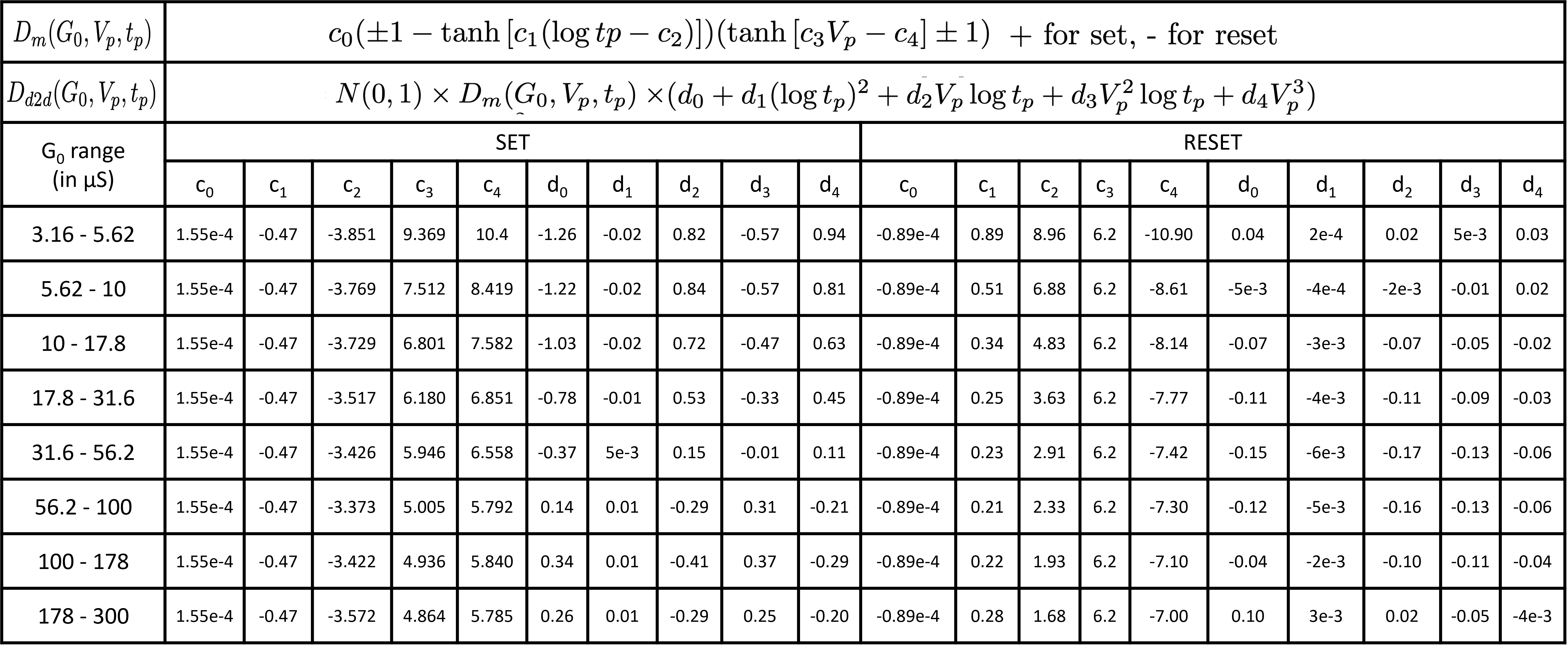}
    \caption{The Dynamic model equations along with the model parameters}
\end{figure*}

\renewcommand{\thefigure}{7}
\begin{figure*}[!b]
    \centering
    \includegraphics[width=1.1\textwidth]{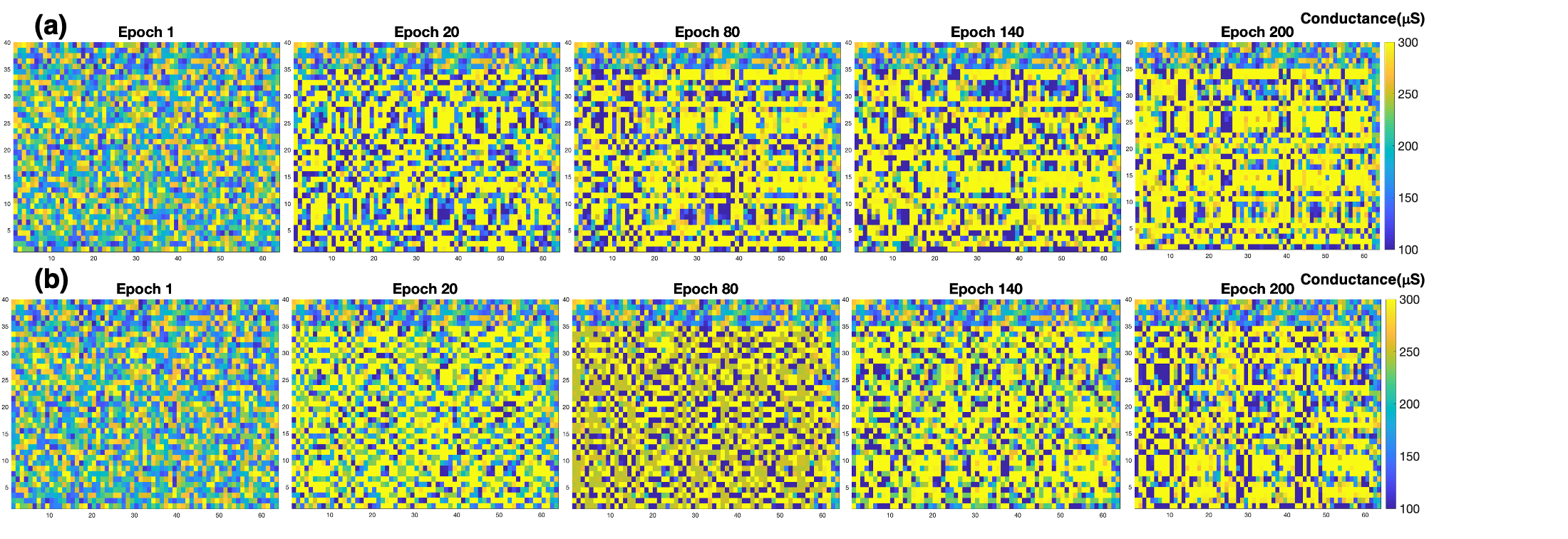}
    \vspace{-25pt}
    \caption{The conductance evolution after different epochs of in-situ training utilising Manhattan's rule for a ($40 \times 64$) (a) variation/noise-free passive crossbar array and (b) passive RRAM crossbar array considering device-to-device variations.}
\end{figure*}

The mean-squared error was used as the loss function during in-situ training and the desired weight change $\Delta W^t$ was calculated using the stochastic gradient descent with momentum in the digital domain, given by:
\begin{align} 
\Delta W^{t} = \alpha \cdot GRAD + \eta \cdot \Delta W^{t-1}
\end{align}
The learning rate $\alpha$ and the momentum value $\eta$ were taken as 0.01 and 0.9, respectively and the conductance-to-weight ratio was selected as $10^{-4}$. Manhattan rule is a coarse-grain variation of backpropagation algorithm where the weight update is based on the sign information of weight gradients. Based on the sign of the desired weight change $\Delta W^{t}$ (obtained using equation (7)), a single voltage pulse of fixed amplitude ($V_{reset}$ = – 0.8 V for depression and $V_{set}$ = 0.8 V for potentiation) and fixed duration (100 ns) was applied to change the RRAM conductance in the appropriate direction as:
\begin{align}
G = G_0 + \Delta G_0
\end{align}

$$
    V_p = \begin{cases}
            V_{set},& \text{if $\Delta W^{t} > 0$}\\
            V_{reset},& \text{if $\Delta W^{t} < 0$}\\
            0,              & \text{otherwise}
           \end{cases}
$$

The conductance change ($\Delta G_0$) of the RRAMs in a passive crossbar array follows the dynamic equation [16]:
\begin{align}
\Delta G_0 = D_{m}(G_0,V_p,t_p) + D_{d2d}(G_0,V_p,t_p)
\end{align}
where $D_{m}$ is the expected noise-free conductance change after application of the voltage pulse (with amplitude $V_p$ and duration $t_p$) which also depends on the present conductance-state leading to non-linearity, and $D_{d2d}$ is the normally-distributed stochastic device-to-device variations for different RRAMs on the same crossbar array.

This fixed amplitude training using the Manhattan's rule is hardware-friendly since it obviates the need for computation and storage of the exact values of the gradient [21]. Furthermore, the calculation of the activation functions ($tanh$ and sigmoid, $\sigma$) on the gate layer outputs of the LSTM cells were performed in the digital domain.
\par For performance benchmarking, the proposed LSTM network is applied to a standard time series prediction (regression) problem similar to [10] where the task is to forecast the number of international airline passengers for the next month based on the prior observations.

\renewcommand{\thefigure}{8}
\begin{figure}[t]
    \centering
    \includegraphics[width=0.48\textwidth]{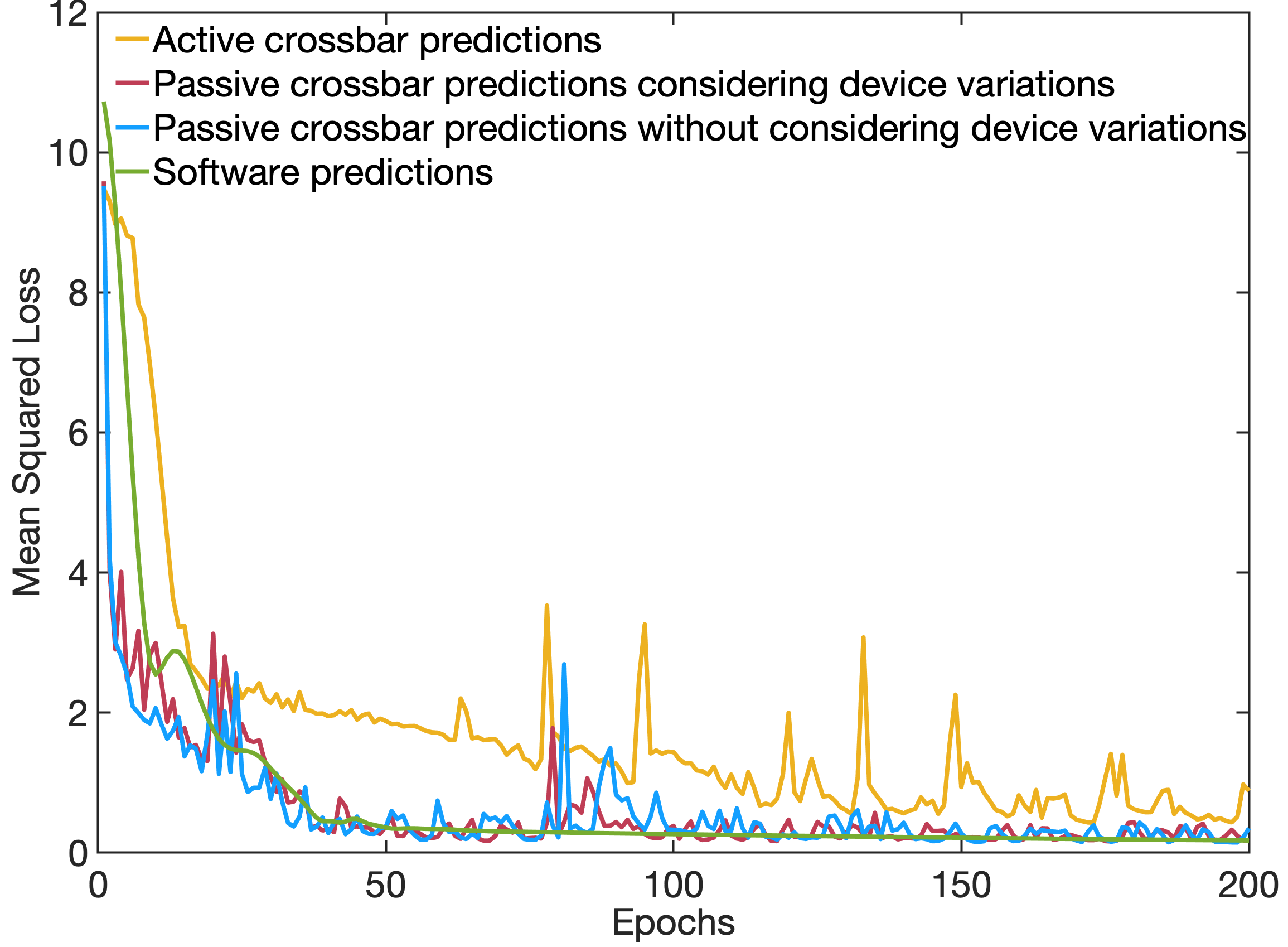}
    \caption{The mean-squared loss function of the different LSTM network implementations during training for 200 epochs.}
\vspace{-7pt}
\end{figure}

The data set consists of the number of passengers from January 1949 to December 1960 with 144 observations. The first 96 observations were used for training and the remaining 48 observations were used for testing. The passenger data was normalized before applying as inputs. One observation ($x^t$) was appended with the hidden cell state $h^{t-1}$ of the previous time step and a bias to form an input vector. The input vector was multiplied by a read voltage $(V_{read})$ and padded by zeros to form a vector of input voltages $(V_{in})$ before applying to the passive RRAM crossbar array based LSTM layer comprising of four gates $(h_a^t, h_i^t, h_o^t, h_f^t)$. The hidden cell state ($h^t$) obtained from the LSTM layer was appended with a bias and multiplied by a read voltage $(V_{read})$ to form the input voltage vector $(V_{in})$ for the dense layer. During the inference, the output column current of the dense layer represents the prediction for the next month and is scaled back to get the passenger count.

\section{Results and Discussion}
Utilizing the hardware-aware simulation framework developed in section II, we performed a comprehensive analysis of the proposed LSTM network implementation exploiting the passive RRAM crossbar array and compared the different performance metrics such as accuracy, area and energy consumption during the training against the software (digital) and active 1T-1R RRAM array-based implementations [10]. To decouple the impact of hardware imperfections such as spatial and temporal variations and noise on the performance of the proposed LSTM-based RNNs, we have also performed simulations considering variation/noise-free passive RRAM crossbar (exhibiting non-linear conductance update behavior) by switching-off the variability and noise flags in the comprehensive compact model [16].

\subsection{Accuracy}
The mean-squared loss function of the recurrent-LSTM network (described in section II) during the training using backpropagation with stochastic gradient descent algorithm for digital (software) implementation, in-situ training using a hybrid of stochastic gradient descent and resilient back propagation for active 1T-1R RRAM crossbar array-based implementation [10] and in-situ training using Manhattan rule for the proposed passive RRAM crossbar array-based implementation are shown in Fig 5. The LSTM model trained over passive RRAM crossbar array shows a faster convergence (within 200 epochs) as compared to the active 1T-1R-based implementation [10]. Therefore, the in-situ training of the LSTM networks on the passive RRAM crossbar arrays leads to a reduced latency. The conductance evolution of the passive RRAM crossbar array during the in-situ training is also shown in Fig. 7. As can be observed from Fig. 7, the conductance map does not change significantly even in the presence of hardware non-idealities such as device-to-device variations and noise.
\par
The passenger-count prediction after 200 epochs of training for digital (software) implementation, active 1T-1R RRAM crossbar array-based implementation [10] and the proposed passive RRAM crossbar array-based implementation (considering the non-idealities) are shown in Fig. 9. It is evident from Fig. 9 that the predictions made by the passive RRAM crossbar-based LSTM implementation follows the observations (test dataset) more closely as compared to their software and active 1T-1R crossbar counterparts. Moreover, Fig. 10 compares the predictions made by the LSTM implementation with variation/noise-free RRAM cossbar and passive crossbar considering the device-to-device variations and noise after in-situ training for 200 epochs. Fig. 10 clearly indicates that the LSTM model considering device-to-device variations and noise also performs well and fits the actual observations with decent accuracy. Therefore, the proposed LSTM network implementation based on the passive RRAM crossbar array is resilient to hardware artifacts such as spatial and temporal variations, noise and non-linearity.
 
 \renewcommand{\thefigure}{9}
\begin{figure}[t]
    \centering
    \includegraphics[width=\linewidth]{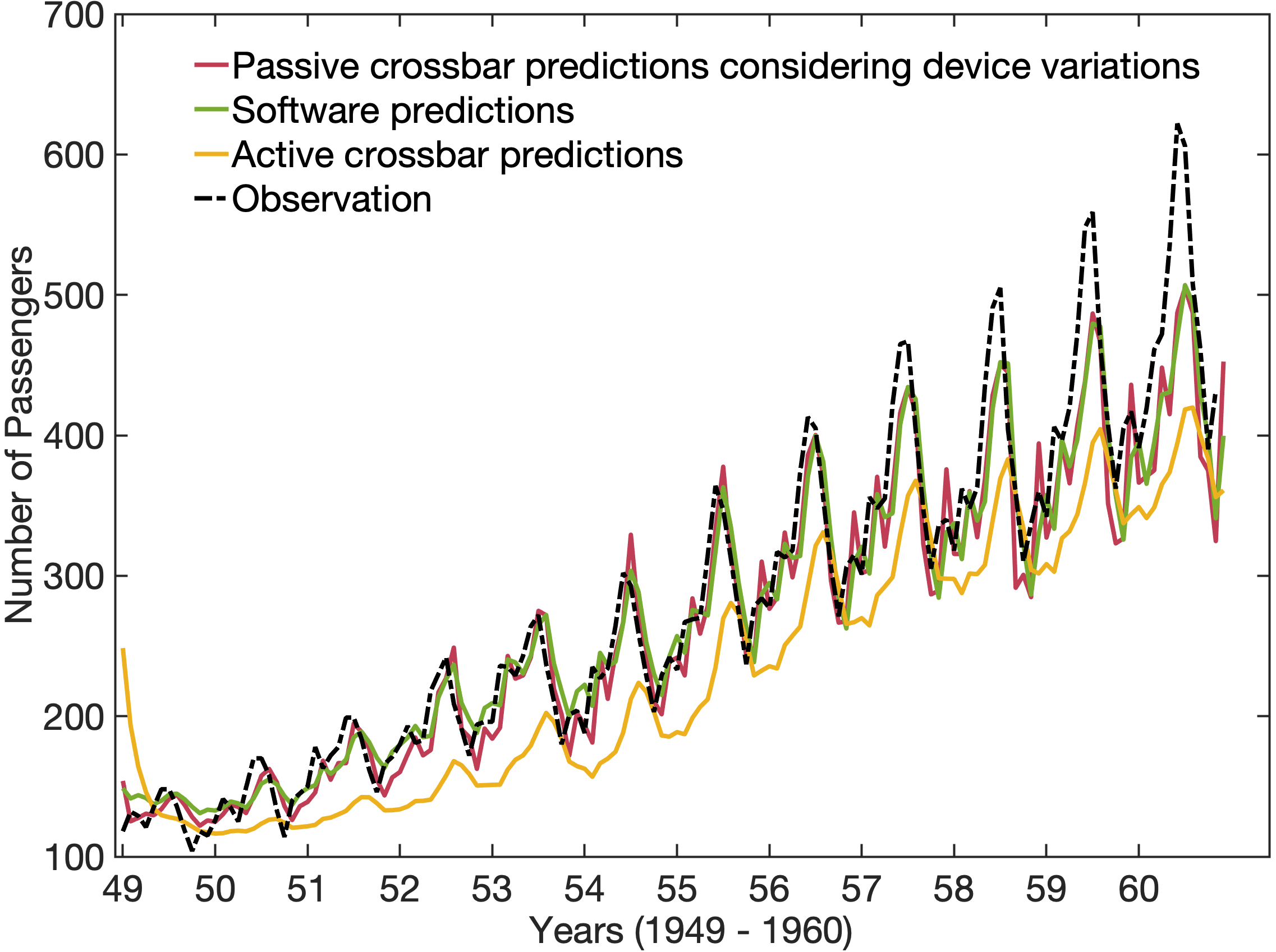}
    \vspace{-10pt}
    \caption{The performance of the different LSTM network implementations after training for 200 epochs in predicting the observations (data set).}
    \vspace{-12pt}
\end{figure}

\renewcommand{\thefigure}{10}
\begin{figure}[t]
    \centering
    \includegraphics[width=\linewidth]{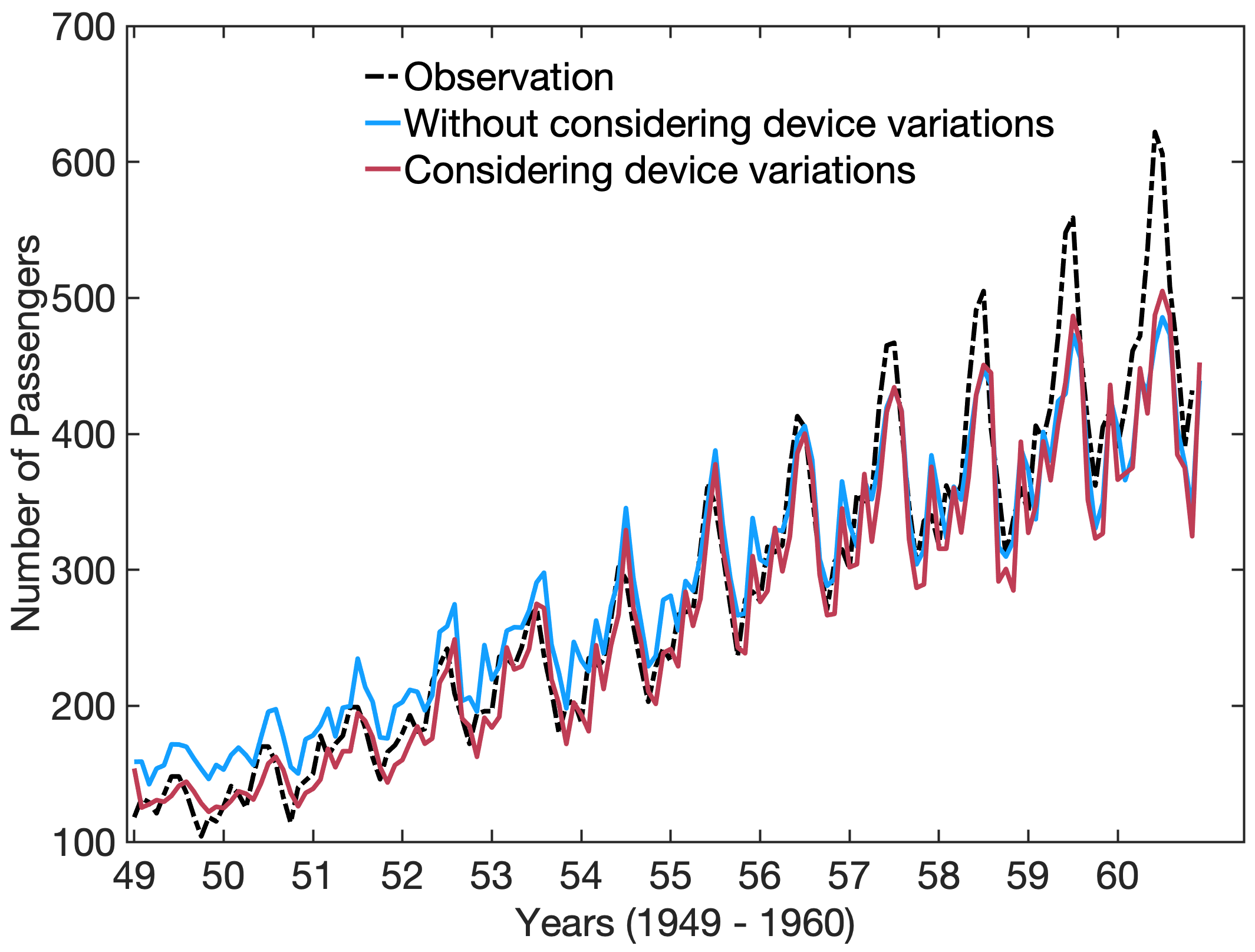}
    \vspace{-10pt}
    \caption{The performance of the LSTM network implementation based on variation/noise-free RRAM crossbar and passive RRAM crossbar array considering hardware non-idealities after in-situ training for 200 epochs in predicting the observations (data set).}
    \vspace{-7pt}
\end{figure}

\subsection{Energy consumption during training}
We also calculated the energy dissipated in the passive RRAM crossbar array during the in-situ training of the LSTM network using Manhattan's rule. The application of the programming (set or reset) pulse to tune the conductance-state ($G_{i-1}$) of the passive RRAM crossbar array leads to an energy consumption ($E_i$) depending on the gradient sign i.e. whether the conductance needs to be potentiated or depressed as:
$$
    E_i = \begin{cases}
            V_{set}^2 \cdot G_{i-1} \cdot t_p,& \text{if $G_{i} > G_{i-1}$}\\
            V_{reset}^2 \cdot G_{i-1} \cdot t_p,& \text{if $G_{i} < G_{i-1}$}\\
            0,              & \text{otherwise}
        \end{cases}
$$

The energy consumed in the passive RRAM crossbar array after each epoch of in-situ training of the proposed implementation is shown in Fig. 11. The cumulative energy required for updating the conductance-states of the variation/noise-free RRAM crossbar and the passive RRAM crossbar array considering the non-idealities are also shown in Fig. 12. The total energy required for in-situ training of the passive RRAM crossbar-based LSTM implementation till the network converges to optimal accuracy (200 epochs) considering the device-to-device variations and noise (3.0 $\mu J$) is somewhat larger than the total energy consumed by variation/noise-free passive RRAM crossbar array (2.8 $\mu J$).
\par Utilizing the same approach, we also calculated the energy consumed while performing conductance-update during in-situ training of the active 1T-1R crossbar-based LSTM implementation [10]. To simplify the energy calculations for the active 1T-1R crossbar array, an average conductance of $500\mu S$ was assumed by the authors in [10] to determine the average energy dissipation for $V_{set} = 2.5 V$ and $V_{reset} = 1.7 V$. The energy dissipated during in-situ training of the network for 800 epochs (till the network converges to optimal accuracy) was found to be $\sim$145 $\mu J$ ($\sim$35 $\mu J$ after 200 epochs of training with sub-optimal accuracy). Therefore, the proposed implementation exhibits a significantly reduced energy dissipation in the RRAM crossbar array by a factor of $\sim$51.7. However, it may be noted that the energy consumed during the gradient calculation in the digital domain was not included in these estimates for both cases. Since the gradient calculation is also easier while using the Manhatten's rule, we believe that the energy consumption during the training process would also follow a similar trend.  

\renewcommand{\thefigure}{11}
\begin{figure}[t]
    \centering
    \includegraphics[width=0.44\textwidth]{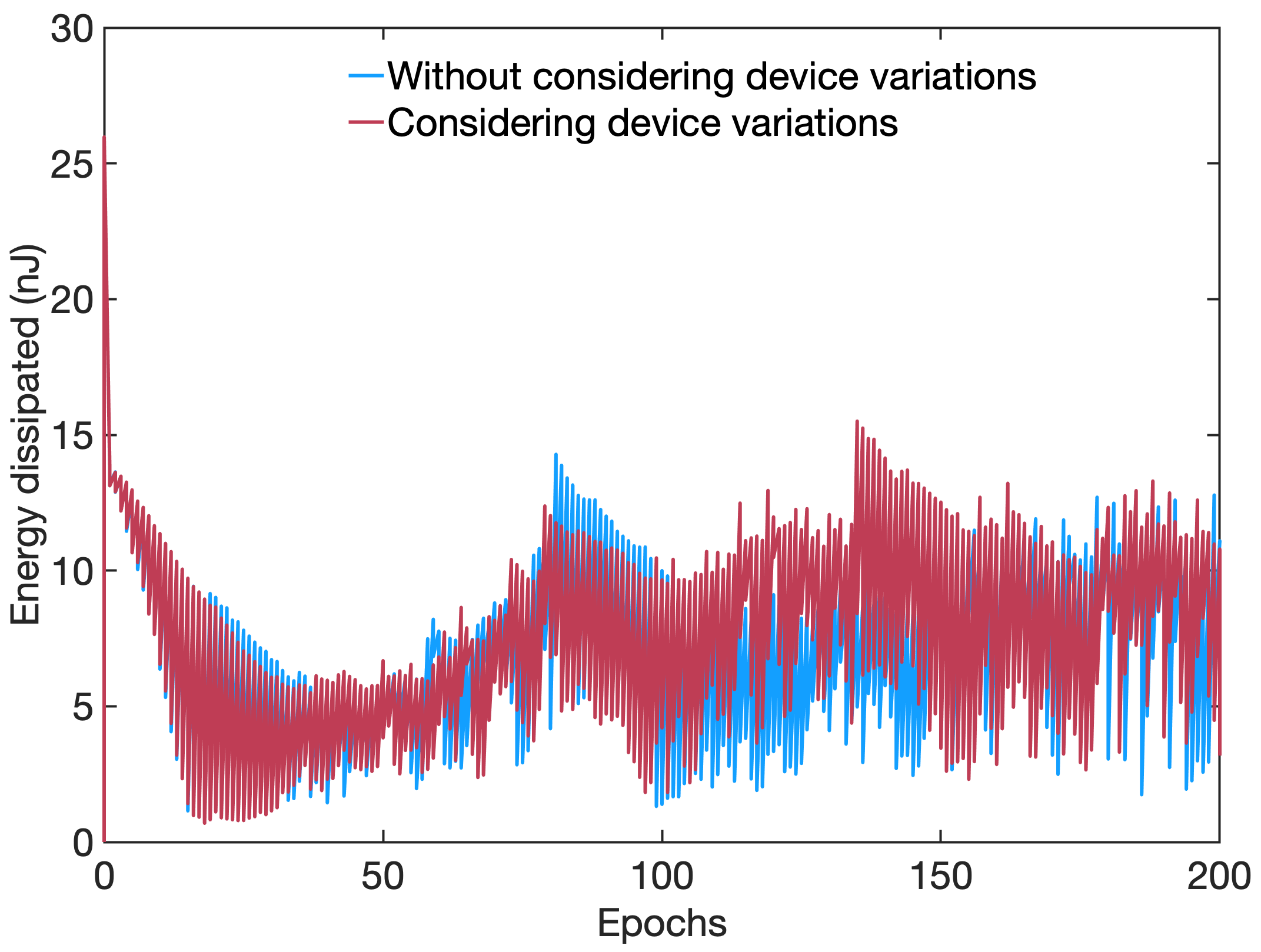}
    \vspace{-5pt}
    \caption{The energy consumed by the proposed passive RRAM crossbar-based LSTM implementation in each epoch during in-situ training.}
    \vspace{-8pt}
\end{figure}

\renewcommand{\thefigure}{12}
\begin{figure}[ht]
    \centering
    \includegraphics[width=0.44\textwidth]{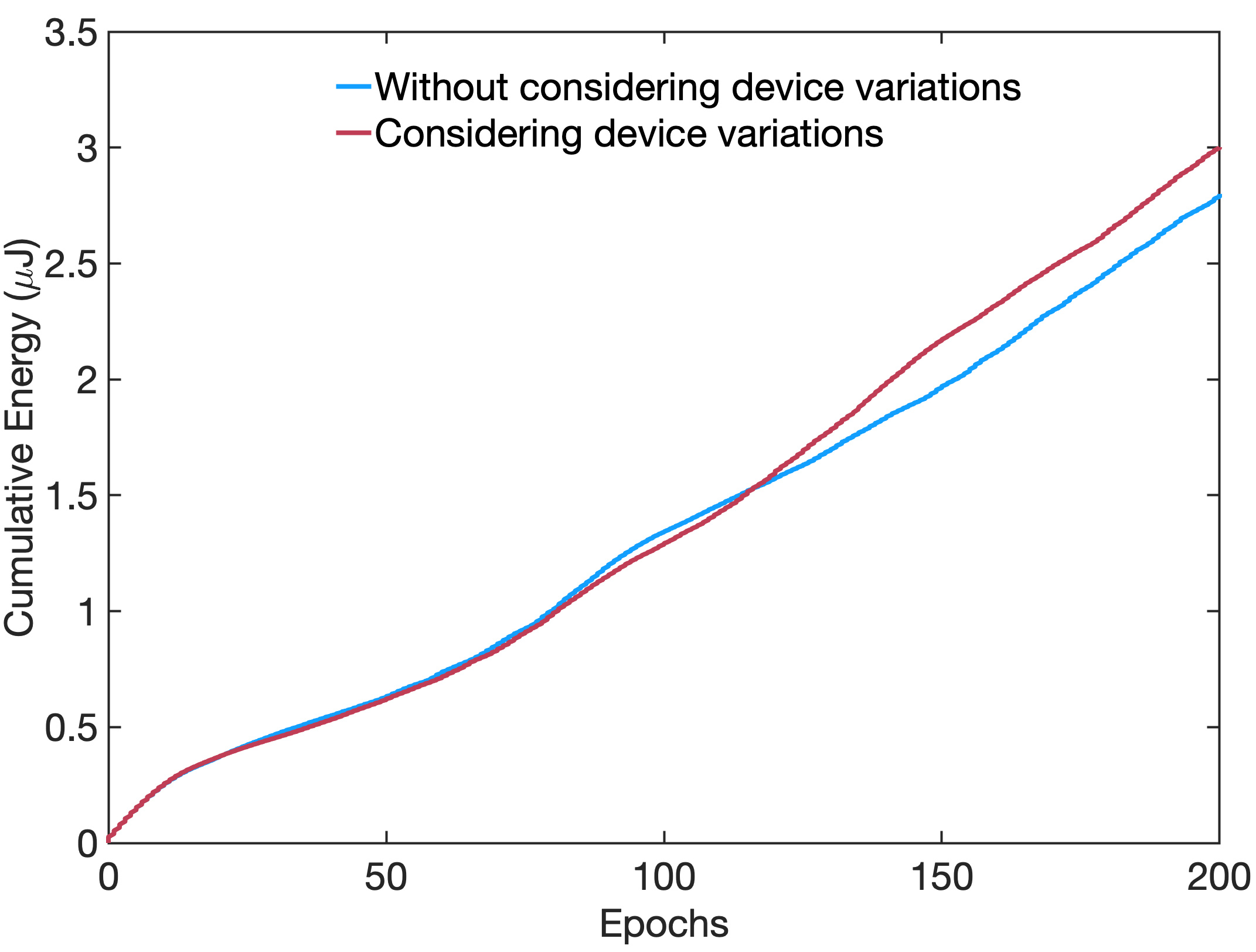}
   \vspace{-5pt}
    \caption{The cumulative energy consumed by the proposed passive RRAM crossbar-based LSTM implementation after each epoch during in-situ training.}
\vspace{-15pt}
\end{figure}

\subsection{Area}
While a single memory cell in the active 1T-1R RRAM array of [10] occupies an area of $\sim$2360 (59 $\times$ 40) $\mu m^2$, the footprint of an RRAM cell in the passive RRAM crossbar [16] is $\sim$0.36 (0.6 $\times$ 0.6) $\mu m^2$ and dictated only by the metal electrode pitch. The overall area required for implementing the core computations of LSTM-based RNN with the active 1T-1R RRAM crossbar array of size 40 $\times$ 64 is approximately 6.041 $mm^2$ [10], while the proposed passive RRAM crossbar array-based implementation requires an area of $\sim$921.6 $\mu m^2$ [16], for a 40 $\times$ 64 array. Therefore, the proposed implementation leads to a significant reduction in the footprint by a factor of $\sim$6.5$\times$10$^3$.

\section{Conclusions}
In this work, for the first time, we propose to perform the resource-intensive core computations of the LSTM network in-situ on a passive RRAM crossbar array for realizing compact and ultra-low power RNN engines for mobile IoT devices. We developed a hardware-aware simulation framework for evaluating the performance of the proposed LSTM implementation utilising an experimentally calibrated comprehensive phenomological model which can be extended for simulation of multi-layer deep LSTM-based RNNs for practical applications. Our extensive investigation reveals that the proposed implementation outperforms the prior digital and active 1T-1R RRAM array-based LSTM implementations by several orders of magnitude in terms of area and energy consumption during the training. Moreover, the area-efficiency of the passive RRAM crossbar-based LSTM networks may be further enhanced by 3D-integration of several RRAM layers utilizing common electrodes in each layer [15]. Our results may provide the incentive for experimental realisation of such a novel computing paradigm for LSTM-based deep RNNs.

\end{document}